\documentclass[twoside,fleqn]{article}
\usepackage{espcrc2}
\usepackage{theorem}
\usepackage{epsfig}
\usepackage{graphicx}

\theoremstyle{break}    \newtheorem{The}{Theorem}
\theoremstyle{plain}    
\theoremstyle{plain}    
\theoremstyle{plain}    
{\theorembodyfont{\rmfamily}     }
{\theorembodyfont{\rmfamily}     }

\def\gsim{{\mathrel{\raise2pt\hbox to 8pt{\raise -5pt\hbox{$\sim$}\hss{$>$}}}}}
\def\rsim{{\mathrel{\raise2pt\hbox to 8pt{\raise -5pt\hbox{$\sim$}\hss{$>$}}}}}
\def\lsim{{\mathrel{\raise2pt\hbox to 8pt{\raise -5pt\hbox{$\sim$}\hss{$<$}}}}}

\begin{document}

\title{
Recent progress in calculating weak matrix elements using staggered
fermions\thanks{Research done in collaboration
with T.~Bhattacharya, G.T.~Fleming, G.~Kilcup, R.~Gupta, and
S.~Sharpe.  Research supported in part by BK21, by the SNU foundation
\& Overhead Research fund, by KRF contract KRF-2002-003-C00033 and
by KOSEF contract R01-2003-000-10229-0.}  }
\author{
        Weonjong~Lee\address{School of Physics, 
		Seoul National University,
		Seoul, 151-747, South Korea}
}
\begin{abstract}
We present a chronological review of the progress in calculating weak
matrix elements using staggered fermions.
We review the perturbative calculation of one-loop matching formula
including both current-current diagrams and penguin diagrams using
improved staggered fermions.
We also present preliminary results of weak matrix elements relevant
to CP violation calculated using the improved (HYP (II)) staggered
fermions.
Since the complete set of matching coefficients at the one-loop level
became available recently, we have constructed lattice operators with
all the $g^2$ corrections included.
The main results include both $\Delta I = 3/2$ and $\Delta I = 1/2$
contributions.
\end{abstract}

\maketitle


%
\section{INTRODUCTION}\label{sec:intr}
Staggered fermions preserve enough chiral symmetry to prevent those
operators of our physical interest from mixing with operators of wrong
chirality and to protect quark mass from additive renormalization.
This chiral symmetry is essential to calculate $\epsilon'/\epsilon$.
Staggered fermions have the advantages over DWF of requiring less
computing time and dynamical simulations are already possible with
relatively light quark masses.
By construction, staggered fermions possess four degenerate flavors
(also called ``tastes'').
They allow large taste changing quark-gluon interactions, which make
the perturbative correction large even at the one loop level.
In addition, four-fermion operator mixing matrix is huge ($65536
\times 65536$), which makes it impractical to determine the matching
coefficients in a completely non-perturbative way.
Another disadvantage is that the unimproved staggered fermion action
and operators receive large perturbative corrections
\cite{ref:wlee:100,ref:sharpe:100} and have large scaling violations
of order $a^2$ \cite{ref:jlqcd:1}.
Both of these disadvantages can be alleviated by improving staggered
fermions using fat links.
It has been shown that taste symmetry breaking in the pion spectrum is
significantly reduced with fat links
\cite{ref:orginos:0,ref:hasenfratz:0}.
The first goal of this staggered $\epsilon'/\epsilon$ project is to
check the results ($\epsilon'/\epsilon < 0$) obtained using quenched
domain wall fermions (DWF) by CP-PACS \cite{ref:CP-PACS:0} and RBC
collaboration \cite{ref:RBC:0}.
The second goal is to extend the calculation to dynamical simulation.
The main goal is to find a window for new physics or to confirm the
standard model, when our numerical results are compared with the
experimental results.
This project has progressed through the following small steps.

\section{PERTURBATION}
Since the penguin diagrams had already been known for the gauge
invariant operators from \cite{ref:sharpe:101}, we calculated the
remaining diagrams of the current-current type \cite{ref:wlee:100}.
This provided a complete set of matching formula for
$\epsilon'/\epsilon$, combined with the existing results for penguin
diagrams \cite{ref:sharpe:101}.
We also observed that the perturbative correction for some of
the operators such as $O_7$, $O_8$ and $O_6$ are so large that 
the uncertainty from the truncated contribution is of order 100\%
for using unimproved staggered fermions. 

\section{NUMERICAL STUDY}
Using unimproved staggered fermions, we performed a numerical study on
$\epsilon'/\epsilon$. 
Using the matching formula given in
\cite{ref:wlee:100,ref:sharpe:101}, we constructed fully one loop
matched gauge invariant operators.
As expected, we found large perturbative corrections for $B_7$ and
$B_6$, so that we can not quote quantitative results for this although
the statistical uncertainties are under control.
We also observed that different quenching transcriptions of
the continuum operators on the lattice (proposed by Golterman
and Pallante \cite{ref:golterman:1} lead to noticeably different
values for $B_6$ \cite{ref:wlee:1}.
This indicates a large quenching uncertainty, which deserves further
investigation.

\section{IMPROVEMENT SCHEME}
Hence, it was essential to reduce the large perturbative correction by
improving the staggered fermions.
Hence, the main goal was to find an improvement scheme which can
reduce the perturbative correction down to 10\% or less.
In order to find the best scheme, we calculated, explicitly, one loop
matching coefficients for various improved staggered actions and
operators \cite{ref:wlee:2}: 1) Fat7, 2) Fat7+Lepage, 3) HYP and 4)
AsqTad-like (Fat7+Lepage+Naik) actions.
We observe that all the above improvement schemes significantly reduce
the size of matching coefficients.
After the second level of mean-field improvement, the HYP and Fat7
links lead to the smallest one-loop corrections \cite{ref:wlee:2}.
Since the HYP fat link reduces the non-perturbative flavor symmetry
breaking more efficiently in the pion spectrum than the Fat7 link,
we adopted the HYP scheme in our numerical study.

\section{HYP/$\overline{\rm Fat7}$ }
We studied further on the HYP link \cite{ref:wlee:3}.
The HYP links possess some universal properties, which are summarized
in the following 5 theorems.
\begin{The}[SU(3) Projection]
Any fat link can be expanded in powers of gauge fields ($A_\mu$).
\begin{eqnarray*}
& & B_\mu = B_\mu^{(1)} + B_\mu^{(2)} + B_\mu^{(3)} + \cdots
\\ & & B_\mu^{(n)} = {\cal O}(A^n)
\end{eqnarray*}
  \begin{enumerate}
  \item The linear term, $B_\mu^{(1)}$ is invariant under SU(3)
  projection.
  \item The quadratic term, $B_\mu^{(2)}$ is antisymmetric in gauge
  fields.
  \end{enumerate}
\label{theorem:su(3)}
\end{The}
\begin{The}[Triviality of renormalization]
  \begin{enumerate}
  \item At one loop level, only the $B_\mu^{(1)}$ term contributes
    to the renormalization of the gauge-invariant staggered fermion
    operators.
  \item At one loop level, the contribution from $ B_\mu^{(n)}$ for
    any $n \geq 2$ vanishes.
  \item At one loop level, the renormalization of the gauge-invariant
    staggered operators can be done by simply replacing the propagator
    of the $A_\mu$ field by that of the $B_\mu^{(1)}$ field.
  \end{enumerate}
  This theorem is true, regardless of details of the smearing
  transformation.
\label{theorem:renorm}
\end{The}
\begin{The}[Multiple SU(3) projections]
  \begin{enumerate}
  \item The linear gauge field term $B_\mu^{(1)}$ in the perturbative
    expansion is universal.
  \item In general, the quadratic terms may be different from one
    another.  But all of them are antisymmetric in gauge fields.
  \item This theorem is true, regardless of the details of smearing.
  \end{enumerate}
\label{theorem:multi-su(3)}
\end{The}

\begin{The}[Uniqueness]
  If we impose the perturbative improvement condition of removing the
  flavor changing interactions on the HYP action,
  the HYP link satisfies the following:
  \begin{enumerate}
  \item The linear term $B_\mu^{(1)}$ in perturbative expansion is
identical to that of the SU(3) projected Fat7 links.
  \item The quadratic term $B_\mu^{(2)}$ is antisymmetric in gauge
fields.
  \end{enumerate}
\label{theorem:unique}
\end{The}

\begin{The}[Equivalence at one loop]
  If we impose the perturbative improvement condition to remove the
  flavor changing interactions, at one loop level,
  \begin{enumerate}
  \item the renormalization of the gauge invariant staggered operators
    is identical between the HYP staggered action and those improved
    staggered actions made of the SU(3) projected Fat7 links
    (often called ``$\overline{\rm Fat7}$''),
  \item the contribution to the one-loop renormalization can be
    obtained by simply replacing the propagator of $A_\mu$ by
    that of $B_\mu^{(1)}$.
  \end{enumerate}
  \label{theorem:equivalence}
\end{The}

The first two theorems were used in \cite{ref:sharpe:100} and
\cite{ref:degrand:0}, although they did not present their derivation.
For derivations of all five theorems and further details, refer to
\cite{ref:wlee:3}.
As a result of these theorems, we can prove that for each Feynman
diagram,
\begin{eqnarray}
    \parallel C_{fat} \parallel \ < \ \parallel C_{thin} \parallel \, .
\end{eqnarray}
Here, $ C_{fat} $ ($ C_{thin} $) represents perturbative corrections
to gauge-invariant staggered operators constructed using SU(3)
projected fat links (thin links).
This inequality is not valid for those fat links without SU(3)
projection.
Hence, this lead to a conclusion that we may view the SU(3) projection
of fat links as a tool of tadpole improvement for the staggered
fermion doublers~\cite{ref:wlee:3}.
We also present alternative choices of constructing fat links to
improve staggered fermions in~\cite{ref:wlee:3}.
The above five theorems make the perturbative calculation simpler for
the HYP scheme, because one can perform the calculation merely by
replacing the thin link propagator with that of the HYP links.
This simplicity is extensively used in calculating the renormalization
constants of the four-fermion operators in the next stage.
\begin{table}[t]
\begin{center}
        \begin{tabular}{cc}
        Operators   & $[P \times P][P \times P]_{II}$ \\ \hline
        NAIVE       & $ 2 \times (111.3 - 2 C_N) $    \\
        HYP/$\overline{\rm Fat7}$     & $ 2 \times (6.925 - 2 C_H) $
        \end{tabular}
\end{center}
\caption{One-loop correction to $({\cal O}_3)_{II}$.}
\label{tab:PPPP}
\vspace{-0.3in}
\end{table}

\section{PERTURBATION FOR HYP/$\overline{\rm Fat7}$ (1)}
We calculated the current-current diagrams to obtain perturbative
matching coefficients for the staggered four-fermion operators
constructed using the HYP/$\overline{\rm Fat7}$ links
\cite{ref:wlee:4}.
In particular, we are interested in the $({\cal O}_3)_{II}$ operator
(we use the same notation as in~\cite{ref:wlee:100}), because this
receives large perturbative corrections ($\approx 1$) in the case of
unimproved staggered fermions.
\begin{eqnarray*}
({\cal O}_3)_{II} =
        2 ( [P \times P][P \times P]
        - [S \times P][S \times P] )_{II}
\end{eqnarray*}
In Tables \ref{tab:PPPP} and \ref{tab:SPSP}, we present values of
one-loop corrections to $({\cal O}_3)_{II}$ constructed using both the
HYP/$\overline{\rm Fat7}$ link and unimproved thin link (denoted as
NAIVE).
Here, $C_N = 13.159$ and $C_H = 1.4051$, which correspond to tadpole
improvement contributions.
The results shows that, by choosing the HYP/$\overline{\rm Fat7}$
scheme, the perturbative corrections are reduced to $\approx$ 10\%
level.
In the case of the HYP/$\overline{\rm Fat7}$ scheme, note that the
size of one-loop correction is already under control even without
tadpole improvement.
For further details, refer to~\cite{ref:wlee:4}.

\begin{table}[t]
\begin{center}
        \begin{tabular}{cc}
        Operators   & $[S \times P][S \times P]_{II}$ \\ \hline
        NAIVE       & $ 2 \times (95.6 - 6 C_N) $     \\
        HYP/$\overline{\rm Fat7}$  & $ 2 \times (19.1 - 6 C_H) $
        \end{tabular}
\end{center}
\caption{One-loop correction to $({\cal O}_3)_{II}$.}
\label{tab:SPSP}
\vspace{-0.3in}
\end{table}

\section{PERTURBATION FOR HYP/$\overline{\rm Fat7}$ (2)}
In order to obtain a complete set of matching formula to convert the
lattice results into continuum values, we need to calculate the
penguin diagrams for the gauge-invariant staggered operators
constructed using various fat links \cite{ref:wlee:5}.
One of the main results regarding the diagonal mixing from penguin
diagrams can be summarized into the following theorem:
\noindent {\bf Theorem 1 (Equivalence)} \\
{\sl At the one loop level, the diagonal mixing coefficients of
penguin diagrams are identical between (a) the unimproved (naive)
staggered operators constructed using the thin links and (b) the
improved staggered operators constructed using the fat links such as
HYP (I), HYP (II), Fat7, Fat7+Lepage, and $\overline{\rm
Fat7}$}.\footnote{Note that AsqTad is NOT included on the list. In
this case, by construction the operators are made of the fat links
which are not the same as those used in the action due to the Naik
term. In addition, the choice of the fat links are open and not
unique.}

The details on the proof of this theorem will be given in
\cite{ref:wlee:5,ref:wlee:1000}.

By construction, gluons carrying a momentum close to $k \sim \pi/a$
are physical in staggered fermions and lead to taste changing
interactions, which is a pure lattice artifact.
In the case of unimproved staggered fermions, it is allowed to mix
with wrong taste ($\ne 1$) and the mixing coefficient is substantial.
In contrast, in the case of improved staggered fermions using fat
links of our interest such as Fat7, $\overline{\rm Fat7}$ and HYP
(II), the off-diagonal mixing with wrong taste vanishes and is absent.
In the case of the improvement using HYP (I) and Fat7 + Lepage, the
off-diagonal mixing with wrong taste is significantly suppressed.
The details of this off-diagonal mixing will be given in
\cite{ref:wlee:1000}.
\begin{figure}[t]
\epsfig{file=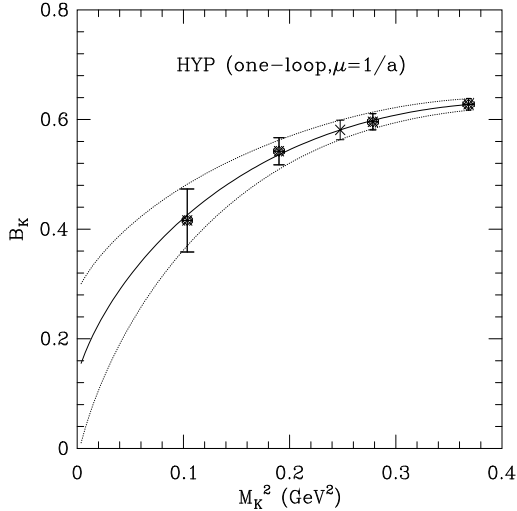, height=16pc, width=16pc}
\vspace*{-5mm}
\caption{$B_K(\mu=1/a)$}
\label{fig:b_k}
\end{figure}
\section{NUMERICAL STUDY FOR HYP/$\overline{\rm Fat7}$}
Recently we have performed a numerical study using Columbia QCDSP
supercomputer with the HYP (II) staggered fermion.
Since we know the complete set of matching formula for
HYP/$\overline{\rm Fat7}$ staggered operators from
\cite{ref:wlee:4,ref:wlee:5}, we constructed fully one-loop matched
gauge-invariant operators to study $\epsilon'/\epsilon$.
Here, we present preliminary estimates of $B_K$, $B_7^{(3/2)}$,
$B_8^{(3/2)}$ and $B_6^{(1/2)}$ calculated using the HYP (II)
staggered fermions at $\beta = 6.0$ on a $16^3 \times 64$ lattice with
218 configurations.
%

%
%

%
\subsection{$B_K$}
\begin{figure}[t]
\epsfig{file=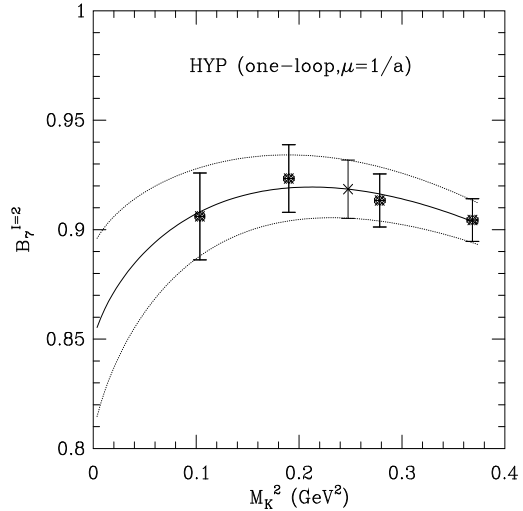, height=16pc, width=16pc}
\vspace*{-5mm}
\caption{$B_7^{\Delta I=3/2}(\mu=1/a)$}
\label{fig:b_7:lin+log}
\end{figure}
\begin{figure}[t]
\epsfig{file=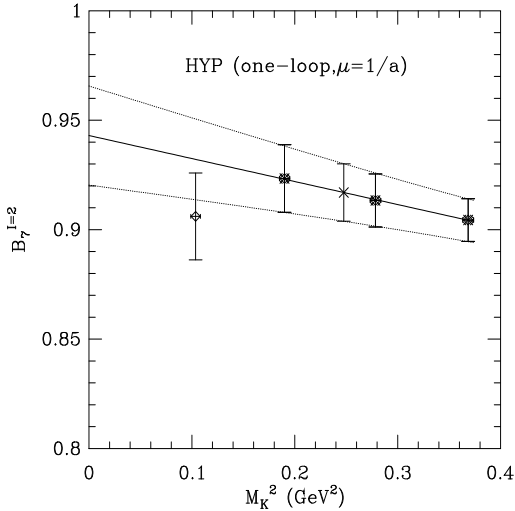, height=16pc, width=16pc}
\vspace*{-5mm}
\caption{$B_7^{\Delta I=3/2}(\mu=1/a)$}
\label{fig:b_7:lin}
\end{figure}
Fig. \ref{fig:b_k} shows $B_K$ as a function of $M_K^2$, where the
mesons are made of degenerate quarks.  
We fit $B_K$ to the form suggested by quenched chiral perturbation
theory \cite{ref:sharpe:1}: $B_K = c_0 + c_1 (M_K)^2 + c_2 (M_K)^2
\log(M_K)^2$.
The cross symbol in Fig. 1 corresponds to the value interpolated to
the physical kaon mass.
Our preliminary result is $B_K = 0.581(18)$, which is consistent with
the continuum extrapolated value calculated using unimproved staggered
fermions \cite{ref:jlqcd:1}.
However, our value is substantially different from the value ($B_K =
0.6790(16)$) calculated using unimproved staggered fermions at finite
lattice spacing ($1/a = 2.01$ GeV).
Therefore, this leads to the preliminary conclusion that, using the
HYP staggered fermions, the scaling behavior is so improved that our
results at $\beta=6.0$ ($1/a = 1.95$ GeV) is already in agreement with
the continuum extrapolated values of the unimproved staggered results.
Of course, this claim needs further investigation at weaker couplings.
In the chiral limit, we obtain $c_0 = 0.13(15)$, which is also
consistent with those results obtained using the NLO, large $N_c$
calculation \cite{ref:peris:1,ref:bijnens:1}.
The value for $c_2/c_0$ is consistent with the predictions of quenched
chiral perturbation theory \cite{ref:sharpe:1} within large errors.
\subsection{$B_7^{(3/2)}$ and $B_8^{(3/2)}$}
\begin{figure}[t]
\epsfig{file=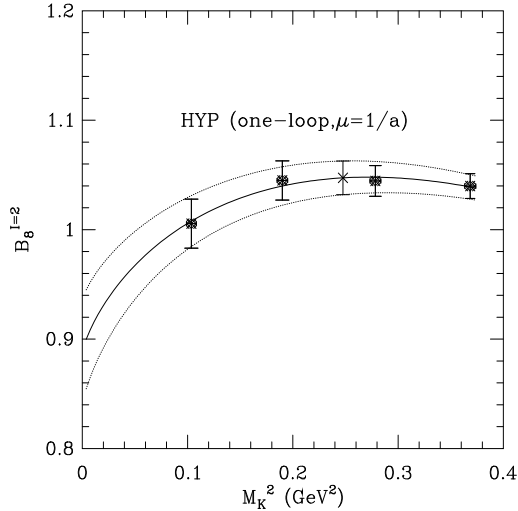, height=16pc, width=16pc}
\vspace*{-5mm}
\caption{$B_8^{\Delta I=3/2}(\mu=1/a)$}
\label{fig:b_8:lin+log}
\end{figure}
\begin{figure}[t]
\epsfig{file=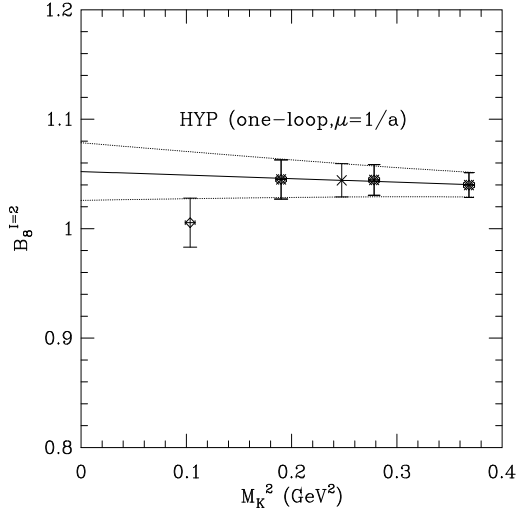, height=16pc, width=16pc}
\vspace*{-5mm}
\caption{$B_8^{\Delta I=3/2}(\mu=1/a)$}
\label{fig:b_8:lin}
\end{figure}
A major contribution to the $\Delta I = 3/2$ amplitudes comes from
$B_7^{(3/2)}$ and $B_8^{(3/2)}$.
We fit $B_7^{(3/2)}$ and $B_8^{(3/2)}$ to the form of
our trial function: $f_1(M_K^2) = c_0 + c_1 (M_K^2) + c_2 (M_K^2)
\log(M_K^2)$.
The results of $B_7^{(3/2)}$ and $B_8^{(3/2)}$ are presented in
Fig.~\ref{fig:b_7:lin+log} and Fig.~\ref{fig:b_8:lin+log},
respectively.
Our preliminary values at the physical kaon mass are
$B_7^{(3/2)}=0.919(13)$ and $B_8^{(3/2)}=1.047(15)$.
We also fit $B_7^{(3/2)}$ and $B_8^{(3/2)}$ to the form suggested by
the quenched chiral perturbation \cite{ref:golterman:1}: $f_2(M_K^2) =
c_0 + c_1 (M_K^2)$.
These fitting results of $B_7^{(3/2)}$ and $B_8^{(3/2)}$ are
shown in Fig.~\ref{fig:b_7:lin} and Fig.~\ref{fig:b_8:lin},
respectively.
Our preliminary values at the physical kaon mass from these fits are 
$B_7^{(3/2)}=0.917(13)$ and $B_8^{(3/2)}=1.044(15)$.
Hence, we observe not only that the interpolated values at the
physical kaon mass are insensitive to the fitting functions, but also
that the extrapolated values in the chiral limit are extremely
sensitive to the fitting functions.
We can think of two possibilities for the chiral extrapolation.
One possibility is that the data point at the lightest quark mass
might be shifted due to a finite volume effect, which certainly need
further investigation in near future.
Another possibility is that the truncated higher order corrections
from the quenched chiral perturbation theory are not negligible so
that we might have to include higher order terms to do better chiral
extrapolation.
For these reasons, we do not quote our values for $B_7^{(3/2)}$ and
$B_8^{(3/2)}$ in the chiral limit but we only quote the values
interpolated to the physical kaon mass as above.
Note that we calculated $B_7^{(3/2)}$ at the scale $\mu = 1/a$
using the HYP staggered fermions, which would not have been meaningful
for unimproved staggered fermions due to large perturbative
corrections.
Compared with previous calculation done using Landau-gauge operators
\cite{ref:kilcup:1}, the systematics of the HYP staggered operators
are significantly reduced and the results are as much more
reliable.
\subsection{$B_6^{(1/2)}$}
\begin{figure}[t]
\epsfig{file=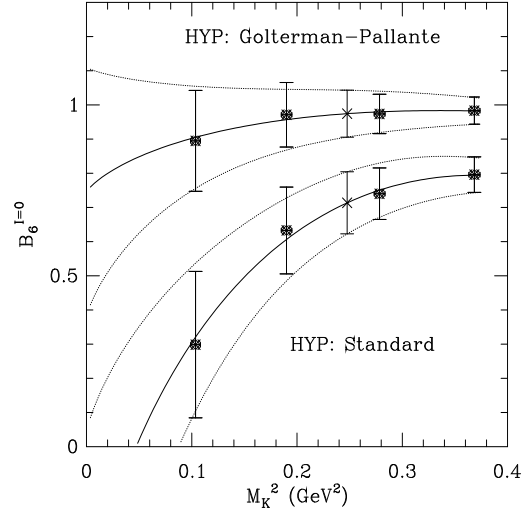, height=16pc, width=16pc}
\vspace*{-5mm}
\caption{$B_6^{\Delta I=1/2}(\mu=1/a)$}
\label{fig:b_6:lin+log}
\end{figure}
\begin{figure}[t]
\epsfig{file=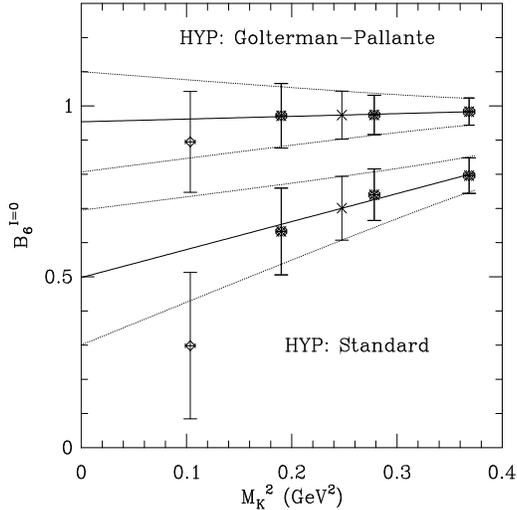, height=16pc, width=16pc}
\vspace*{-5mm}
\caption{$B_6^{\Delta I=1/2}(\mu=1/a)$}
\label{fig:b_6:lin}
\end{figure}

%
%
%
A major contribution to $\Delta I = 1/2$ amplitudes comes from
$B_6^{(1/2)}$.
There are two independent methods to calculate $B_6^{(1/2)}$ in
(partially) quenched QCD: the standard (STD) method and the
Golterman-Pallante (GP) method \cite{ref:golterman:1}.
First, we fit $B_6^{(1/2)}$ to the form suggested by the quenched
chiral perturbation theory \cite{ref:golterman:1}: $f_1(M_K^2) = c_0 +
c_1 (M_K^2) + c_2 (M_K^2) \log(M_K^2)$.
The results are presented in Fig.~\ref{fig:b_6:lin+log}.
Our preliminary values at the physical kaon mass are
\begin{eqnarray*}
B_6^{(1/2),STD}(\mu=1/a) &=& 0.714(91) \\
B_6^{(1/2),GP}(\mu=1/a) &=& 0.974(69) \,.
\end{eqnarray*}
where $1/a = 1.95$ GeV, set by the $\rho$ meson mass.
We also fit $B_6^{(1/2)}$ to another form suggested by the quenched
chiral perturbation \cite{ref:golterman:1}: $f_2(M_K^2) = c_0 + c_1
(M_K^2)$.
The results are presented in Fig.~\ref{fig:b_6:lin}.
Our preliminary values at the physical kaon mass from this fit are 
\begin{eqnarray*}
B_6^{(1/2),STD}(\mu=1/a) &=& 0.701(93) \\
B_6^{(1/2),GP}(\mu=1/a) &=& 0.973(70) \,.
\end{eqnarray*}
As in the case of $B_7^{(3/2)}$ and $B_8^{(3/2)}$, we observe that the
extrapolated values in the chiral limit are extremely sensitive to the
fitting function choice.
At present, we can not exclude the possibility that the data point at
the lightest quark mass could be shifted due to a finite volume effect
(in the $\epsilon$ region of the chiral perturbation theory).
It is also possible that truncated higher order corrections from the
quenched chiral perturbation theory are so significant that we might
have to include these neglected terms for chiral extrapolation.
Therefore, we cannot quote our values of $B_6^{1/2}$ in the chiral
limit.
Note that unlike the unimproved staggered fermion calculations where
the perturbative corrections to the STD calculation are $\approx
50\%$, the perturbative corrections in this calculation are modest for
both the STD and GP methods.
We also observe that the gap between the STD and GP methods is
reduced at the physical kaon mass using the HYP staggered fermions
compared that of the unimproved staggered fermions.
\subsection{Preliminary $\epsilon'/\epsilon$}
We use the formula provided in \cite{ref:buras:1} to convert $(m_s +
m_d)_{\mu=m_c}$, $B_6^{(1/2)}(\mu=m_c)$ and $B_8^{(3/2)}(\mu=m_c)$
at $m_c = 1.3$ GeV into $\epsilon'/\epsilon$.
As mentioned before, we have two different methods to calculate
$B_6^{(1/2)}$ in quenched QCD: the STD and GP methods.
When we use the STD method for $B_6^{(1/2)}(\mu = m_c)$, we obtain
$\epsilon'/\epsilon (STD) = 0.00046(23)$.
For the GP method for $B_6^{(1/2)}(\mu=m_c)$, we obtain
$\epsilon'/\epsilon (GP) = 0.00115(17)$.
These values are very preliminary and we have not included an analysis
of the systematic errors.
In addition, we did not use lattice values for any $B_i^{(1/2)}$
except for $B_6^{(1/2)}$ since we have not yet extracted them.

\section{FUTURE PERSPECTIVES}
We plan to calculate all the $B_{i}^{(1/2)}$ and incorporate
all of them into the calculation of $\epsilon'/\epsilon$.
We also plan to obtain the optimal matching scale, $q^*$
\cite{ref:wlee:6}.
We plan to extend our calculation to dynamical simulation.
Current calculation of $\epsilon'/\epsilon$ uses the leading order
chiral perturbation results to convert $K \rightarrow \pi$ and $K
\rightarrow 0$ amplitudes into $K \rightarrow \pi\pi$ amplitudes.
In principle, it is possible to calculate $K \rightarrow \pi\pi$
amplitudes directly on the lattice.
A number of attempts in this direction have been tried
\cite{ref:norman:1,ref:ishizuka:1}.
There has been a numerical study on $\overline{\rm Fat7}$ as a new fat
link computationally cheaper than the HYP link, while preserving all
the nice features of the HYP link \cite{ref:wlee:2000}.

\section{ACKNOWLEDGMENT}
We thank N.~Christ, C.~Jung, C.~Kim, G.~Liu, R.~Mawhinney and L.~Wu
for their support of this project and assistance with numerical
simulations on the Columbia QCDSP supercomputer.
Helpful discussion with A.~Soni and A.~Hasenfratz is acknowledged
with gratitude.
%
%
%

%
%
%
%

\end{document}